\documentclass[11pt,preprint]{emulateapj-rtx4}
\slugcomment{2014, ApJ, 784, 50}
\usepackage[colorlinks,
            linkcolor=red,
            anchorcolor=blue,
            citecolor=blue
            ]{hyperref}

\shorttitle{Imaging and Spectroscopic Observations of a Filament Channel}
\shortauthors{Chen, Harra, \& Fang}

\begin{document}

\title{Imaging and Spectroscopic Observations of a Filament Channel and the
	Implications for the Nature of Counter-streamings}
\author{P. F. Chen\altaffilmark{1,2}, L. K. Harra\altaffilmark{3}, and C. Fang\altaffilmark{1,2}}
\vspace{0.5cm}
\affil{$^1$ School of Astronomy \& Space Science, Nanjing University,
	Nanjing 210093,	China; \email{chenpf@nju.edu.cn}}
\affil{$^2$ Key Lab of Modern Astron. \& Astrophys. (Ministry of
	Education), Nanjing University, China}
\affil{$^3$ UCL-Mullard Space Science Laboratory, Holmbury St. Mary,
        Dorking RH5 6NT, UK}

\begin{abstract}
The dynamics of a filament channel are observed with imaging and spectroscopic telescopes before and during the filament eruption on 2011 January 29. The extreme ultraviolet (EUV) spectral observations reveal that there is no EUV counterparts of the H$\alpha$ counter-streamings in the filament channel, implying that the ubiquitous H$\alpha$ counter-streamings found by previous research are mainly due to longitudinal oscillations of filament threads, which are not in phase between each other. However, there exist larger-scale patchy counter-streamings in EUV along the filament channel from one polarity to the other, implying that there is another component of uni-directional flow (in the range of $\pm$10 km s$^{-1}$) inside each filament thread in addition to the implied longitudinal oscillation. Our results suggest that the flow direction of the larger-scale patchy counter-streaming plasma  in the EUV is related to the intensity of the plage or active network, with the upflows being located at brighter areas of the plage and downflows at the weaker areas. Besides, we propose a new method to determine the chirality of an erupting filament based on the skewness of the conjugate filament drainage sites. It suggests that the right-skewed drainage corresponds to sinistral chirality, whereas the left-skewed drainage corresponds to dextral chirality.
\end{abstract}


\section{Introduction}\label{sect:intro}

Filaments are conspicuous dark strip-like structures on the solar disk,
especially in chromospheric lines such as H$\alpha$. When viewed above the
solar limb, they are seen to be bright and suspended in the corona, sometimes
with several feet connecting to the solar surface \citep{hira85}. 
High-resolution observations indicate that the strip-like filaments consist of
a collection of thin threads, which make a small angle with the magnetic 
polarity inversion line (PIL) in the photosphere below \citep{tand95}. Assuming that the threads are aligned with the local magnetic field, it is deduced that the magnetic field hosting the filaments is strongly sheared or even twisted. Such a pattern is strongly indicative of non-potentiality of the magnetic field, which could power coronal mass ejections (CMEs) and solar flares. That is why CMEs and big flares are often associated with filament eruptions \citep{chen11, shib11, webb12}. Therefore, it might be fair to say that filaments, when erupting, are not only the core of CMEs, but also are the core of CME research. There are a variety of intriguing features as well as puzzles in both the morphology and dynamics of filaments.

Regarding the morphology, filament threads are not randomly aligned with respect to the local PIL. They are lined up along the magnetic PIL. When viewed from above, some threads extend further away from the spine of the filament, forming a structure called ``barbs". Similar to the ramps of an elevated highway, barbs can be divided into left-bearing and right-bearing types. Most filaments in the northern hemisphere have
right-bearing barbs, and are overlaid by left-skewed coronal arcades, which
have negative magnetic helicity. However, most filaments in the southern
hemisphere have left-bearing barbs, and are overlaid by right-skewed coronal
arcades, which have positive helicity \citep{mart98}. This chirality pattern
is consistent with the hemispheric rule for the magnetic helicity of 
active regions \citep{pevt95, bao98}, hence provides strong constraints on the
modeling of the magnetic structure of the filament channel, as well as its
formation mechanism \citep[see][for a review]{mack10}. In some models, the
magnetic field lines of a filament are sheared arcades running from the
positive polarity of the photosphere directly to the negative polarity side,
either with a dip \citep{anti94} or without a dip \citep{mart94} above the PIL.
In other models, the magnetic field of a filament contains a flux rope, where
the field lines cross the PIL at least 3 times when projected to the solar 
disk \citep[e.g.,][]{aula98}. The application of these models to observations needs to be clarified.

Regarding the dynamics, the plasma in the filament threads is never static, 
even when the whole filament looks invariant. There are both transverse motions
where plasma moves with the magnetic field lines and field-aligned motions
\citep{lin03}. One intriguing field-aligned motion is the counter-streaming
observed in H$\alpha$ \citep{zirk98}.
A natural explanation is that the counter-streaming is due to thread
oscillations \citep{lin03}, as simulated by \citet{anti00} and \citet{xia11}.
Another possibility, which was seldom mentioned, is that each thread has a
persistent forward or backward velocity along the magnetic field line, and neighboring threads have opposite moving directions. In this scenario, along the magnetic field of a thread, the plasma from one chromospheric footpoint evaporates into the corona, and then cools down when penetrating into the H$\alpha$ filament body. After running through the filament thread, the plasma is heated by radiation, and becomes coronal plasma until it drains down at the other chromospheric footpoint of the field line. Such a dynamic equilibrium has been analytically studied by \citet{tsin92}, \citet{delz96}, \citet{low05}, and \citet{petr05}. The dynamics might be maintained by the chromospheric heating, whose asymmetry at the two footpoints of a flux tube determines the direction of the plasma motion. If the chromospheric heating is stochastic, e.g., being affected by convective motions in the photosphere, it is expected to result in counter-streamings of H$\alpha$ filament threads. If this mechanism works, we expect to see interweaving blue and red Doppler shifts along the filament channel on each side of the magnetic PIL.

On 2011 January 29, a filament and its channel were observed by both imaging and EUV spectroscopic instruments. This event provides a unique chance to shed light on both the chirality of the filament and the nature of counter-streamings. The observations are described in \S\ref{sec2}, the results are presented in \S\ref{sec3}, and the implications for the magnetic structure of the filament and the nature of counter-streamings are discussed in \S\ref{sec4}, which is then followed by a short summary.

\section{Observations and Data Analysis}\label{sec2}

\begin{figure}
\epsscale{0.7}
\plotone{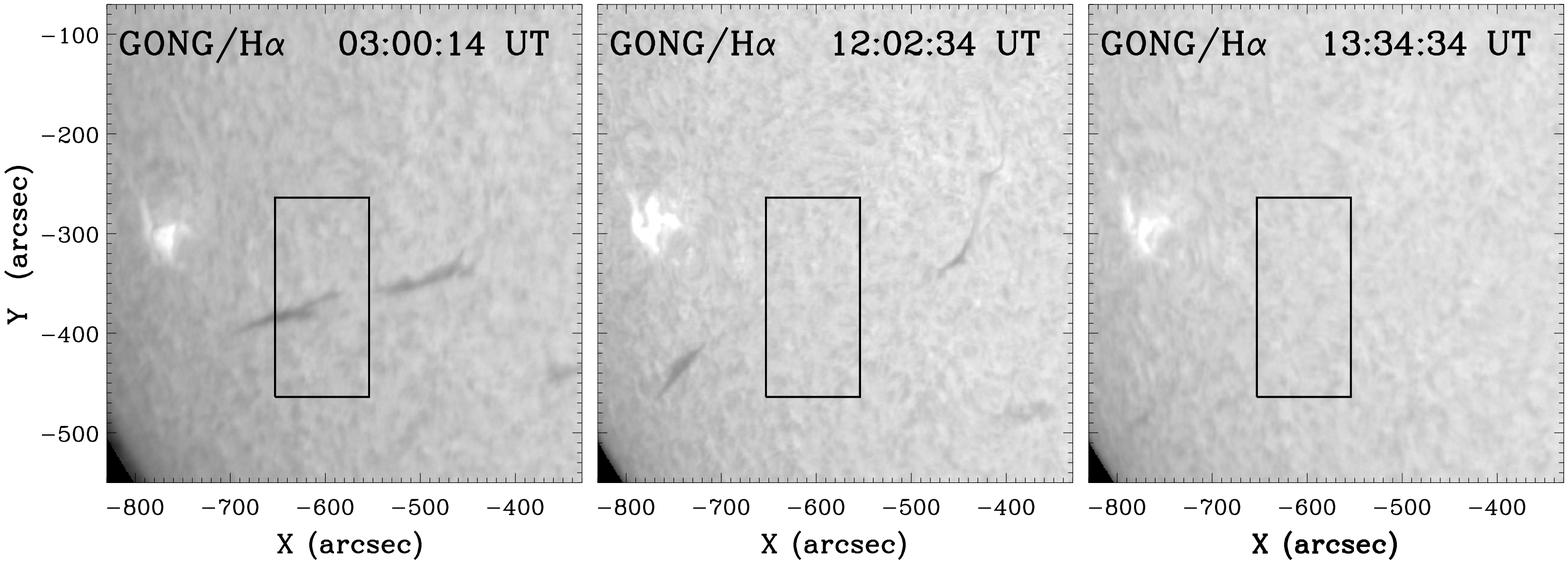}
\caption{Sequential H$\alpha$ images of the filament eruption event on 2011 January 29 event observed by {\it GONG} network. The black boxes mark the field of view of the {\it Hinode}/EIS spectrometer.}
\label{fig1}
\end{figure}

On 2011 January 29, a filament with two segments, as depicted by Figure \ref{fig1}, was located above the magnetic PIL of an extended bipolar magnetic region not far from the active region NOAA 11150, with the Solar Object Locater of SOL2011-01-29T08:00:00L183C110. The filament became activated and started to rise slowly at $\sim$11:00 UT. A weak flaring event appeared from 13:21 UT, with two EUV ribbons and loops visible for more than 10 hrs. The brightening was so weak that no signature was present in the H$\alpha$ images. Correspondingly, there was no enhancement in the {\em GOES} 1--8 \AA\ light curve. 

The event was observed by the EUV Imaging Spectrometer (EIS) on board the {\it Hinode} mission. The {\it Hinode}/EIS \citep{culh07} is a scanning slit spectrometer observing in two wave bands in the EUV: 170--210 \AA\ and 250--290 \AA. The spectral resolution is 0.0223 \AA\ pixel$^{-1}$, which allows velocity measurements of a few km s$^{-1}$. The EIS observation of the event started at 12:02:37 UT and ended at 17:49:42 UT using a sparse raster. This used the  2\arcsec\ slit that rastered across the  filament with a step of 10\arcsec\ and an exposure time of 60 s taking $\sim$11 min for each raster. The field of view (FOV) is 100\arcsec\ in the raster direction and 200\arcsec\ in the slit direction. The EIS observation covered part of the filament and the neighboring filament channel. It thus provides an excellent opportunity to conduct a spectroscopic study of the dynamics of the filament channel. In this study we focus on the coronal line \ion{Fe}{12} 195.12 \AA\ ($\log T$ = 6.1). The spectral profiles are fit with single Gaussian functions. The reference wavelength used in order to determine the velocity was taken from the average of each raster. 

The filament eruption event was also monitored with routine observations from the Global Oscillation Network Group ({\it GONG}) in H$\alpha$ \citep{harv11},  the Atmospheric Imaging Assembly \citep[AIA,][]{leme12} on the {\em Solar Dynamics Observatory} (SDO), and the Extreme Ultra Violet Imager 
\cite[EUVI,][]{howa08} on board the Solar Terrestrial Relations Observatory Behind ({\it STEREO}-B) satellite. The {\it SDO}/AIA observes the Sun in seven EUV and three UV channels with a pixel size of $0\farcs 6$ and a high time cadence of 12 s. In this study we use the 304 \AA, 1600 \AA, and 193 \AA\ bandpasses in order to check the responses from the chromosphere, transition region, and corona. The {\it STEREO}/EUVI observes the Sun with four EUV channels, among which we use the 195 \AA\ band with a pixel size of $1\farcs 59$ and a cadence of 10 min. On 2011 January 29, the {\it STEREO}-B spacecraft was separated from the Earth by an angle of 92.9$^\circ$.

In order to trace the evolution of the filament eruption event, all the images are de-rotated to the universal time 12:02:37 UT when the {\em Hinode}/EIS started the observation. The co-alignment between EIS and AIA is conducted by comparing the locations of a coronal bright point inside the filament channel, which is double-checked by the moving correlation analysis between EIS 195 \AA\ intensity map and AIA 193 \AA\ image. The error of the co-alignment is $\sim$2\arcsec.

\section{Results}\label{sec3}

\subsection{Imaging Observations: from Onset to Eruption}\label{sect:31}

\begin{figure}
\epsscale{.7}
\plotone{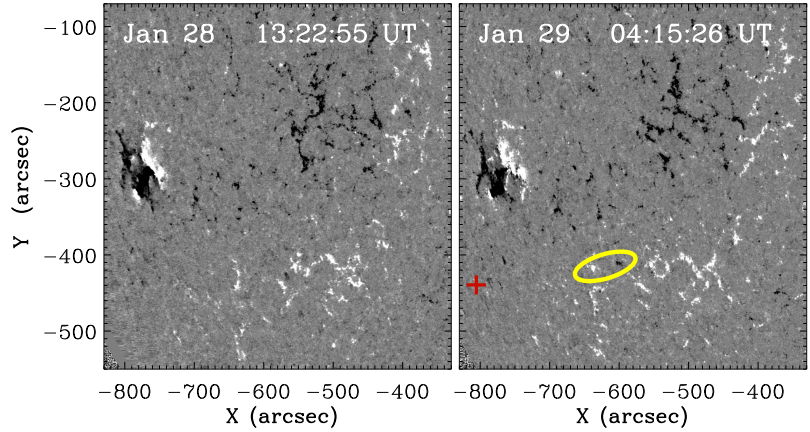}
\caption{Time evolution of the line-of-sight magnetogram observed by {\it SDO}/HMI showing an extended bipolar region where a filament is located above the magnetic polarity inversion line. The yellow ellipse indicates the emergence of small bipolar region, and the red plus sign marks the location of the filament drainage.}
\label{fig2}
\end{figure}

\begin{figure}
\epsscale{0.7}
\plotone{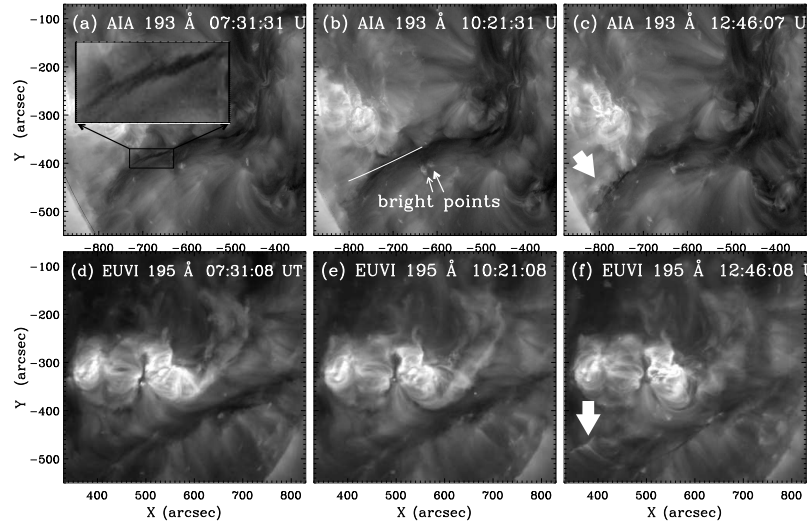}
\caption{Time evolution of the filament activation observed by the {\it SDO}/AIA 193 \AA\ band (top panels) and the {\it STEREO}/EUVI 195 \AA\ band. The inset in panel (a) presents a closeup of a segment of the filament, where left-bearing barbs can be seen clearly. In panel (b), the white line marks the slice in order to plot the time-slice diagram in Fig. \ref{fig4}, and the two thin arrows mark two coronal bright points. The two thick white arrows in panels (c) and (f) mark the locations of the brightenings when the draining filament impacted the solar surface.}
\label{fig3}
\end{figure}

On 2011 January 28, one day before the filament eruption, a small bipolar magnetic structure emerged near the PIL inside the filament channel, as indicated by the yellow ellipse in Figure \ref{fig2}, which shows the evolution of the line-of-sight magnetogram around the filament channel. The extended bipolar region has the normal polarity orientation of the southern hemisphere in solar cycle 24, i.e., positive in the leading polarity and negative in the following polarity. The bipolar flux emergence resulted in two bright points in the EUV images as revealed by Figure \ref{fig3}(b). The polarity orientation of the emerging flux favors magnetic reconnection with the pre-existing magnetic field of the filament channel. Therefore, according to the flux emergence trigger mechanism of CMEs \citep{chen00, kusa12}, the reconnection between the emerging flux and the pre-existing magnetic field relaxes the constraint of the magnetic field over the filament, which then begins to rise.

\begin{figure}
\epsscale{0.7}
\plotone{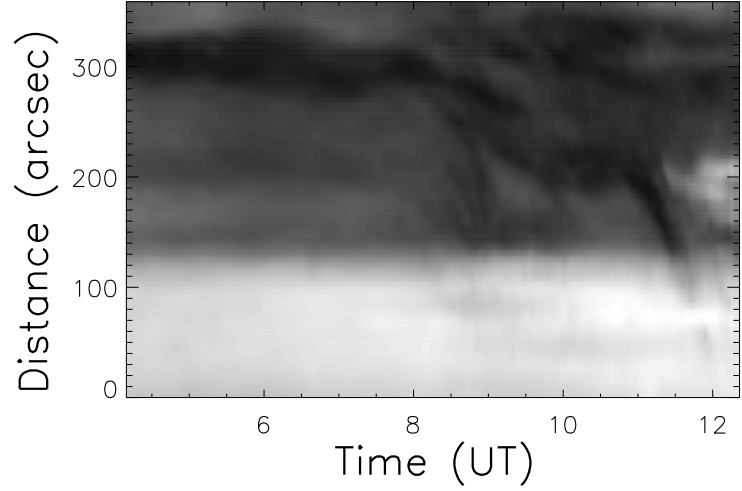}
\caption{Time-distance diagram of the AIA 193 \AA\ intensity along the slice indicated in Fig. \ref{fig3}(b). The start point of the slice is at the southeast end.
}
\label{fig4}
\end{figure}

The filament activation is shown in Figure \ref{fig3}, where the top panels show the time sequence of the {\it SDO}/AIA 193 \AA\ images, and the bottom panels show the {\it STEREO}-B/EUVI 195 \AA\ images at nearly the same time as the top panels. Two bright points consisting of tiny loops, as pointed out in panel (b), were associated with the emerging magnetic flux. The two sets of bright tiny loops are the natural result of the magnetic reconnection between the emerging flux and the overlying antiparallel magnetic field, as simulated by \citet{chen00}. The prominent feature seen from the EUV images and the corresponding animation is that the two segments of the filament, as revealed by the H$\alpha$ images in Figure \ref{fig1}, started to move apart along the filament spine direction from 08:00 UT. In order to see the motion clearly, we select a slice along the east segment of the filament, which is along the great circle of the solar surface as marked by the white line in Figure \ref{fig3}(b). The corresponding time-slice diagram of the AIA 193 \AA\ intensity is depicted in Figure \ref{fig4}, where the start point of the slice is the southeastern end. It is seen that from 08:00 UT to 12:00 UT, the filament material in the selected segment moved eastward, and the velocity was accelerating, partly due to the projection effects since the filament was above the solar surface. At $\sim$11:10 UT, i.e., $\sim$3 hrs after the drainage started, we can see a significant lift-up of the filament from the EUVI movie. It is possible that the mass drainage plays a role in facilitating the onset of the filament eruption \citep{zhou06}.

As the cold filament plasma drained down and impacted the chromosphere, the significant feature in Figure \ref{fig3} is the EUV brightening as pointed by the two thick arrows. The position of the brightening in the AIA 193 \AA\ image at 12:46:47 UT which is shown in the magnetogram in Figure \ref{fig2} as a plus sign. It is revealed that the brightening was located at the negative polarity, to the left of the PIL.

\begin{figure}
\epsscale{0.7}
\plotone{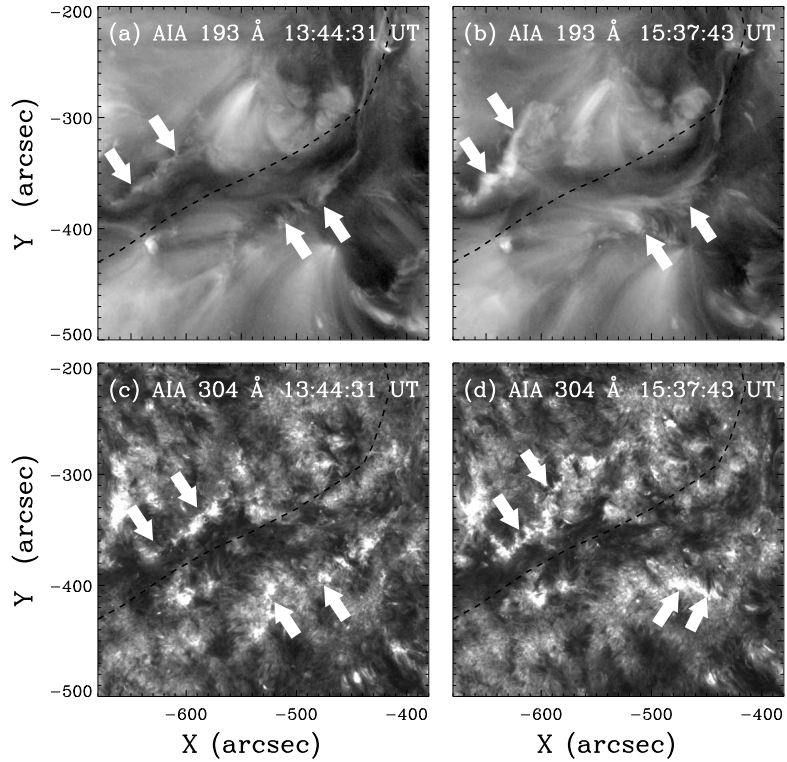}
\caption{Time evolution of the flare ribbons and the flaring arcades observed by AIA 193 \AA\ band (top panels) and AIA 304 \AA\ band (bottom panels). The flare ribbons are marked by the white arrows, and the flaring arcades are visible to connect the two ribbons in panel (b). The flaring arcades are right-skewed compared to the underlying magnetic polarity inversion line, which is marked by the dashed lines.}
\label{fig5}
\end{figure}

After the two filament segments moved apart, the cold plasma near the center of the filament spine was evacuated. As a result, unfortunately, the rise motion of the filament-hosting magnetic field lines cannot be directly observed. About $\sim$30 min after the main draining body of the filament impacted the chromosphere which produced the EUV brightenings, two EUV flaring ribbons appeared in the AIA 193 \AA\ and 304 \AA\ images in Figure \ref{fig5}, where the black dashed lines are the magnetic PIL in the projected plane. Both ribbons were evident in the 304 \AA\ band, but the west ribbon was much weaker than the east ribbon in 193 \AA\ band. The AIA 193 \AA\ image in Figure \ref{fig5}(b) further indicates that a collection of flaring loops connected the two ribbons, overlying the magnetic PIL. Comparing the orientation of the flaring arcade with the PIL, it is revealed that the flaring arcade was right-skewed.

\subsection{Spectroscopic Observation of the Filament Eruption}\label{sect:32}

\begin{figure}
\epsscale{0.7}
\plotone{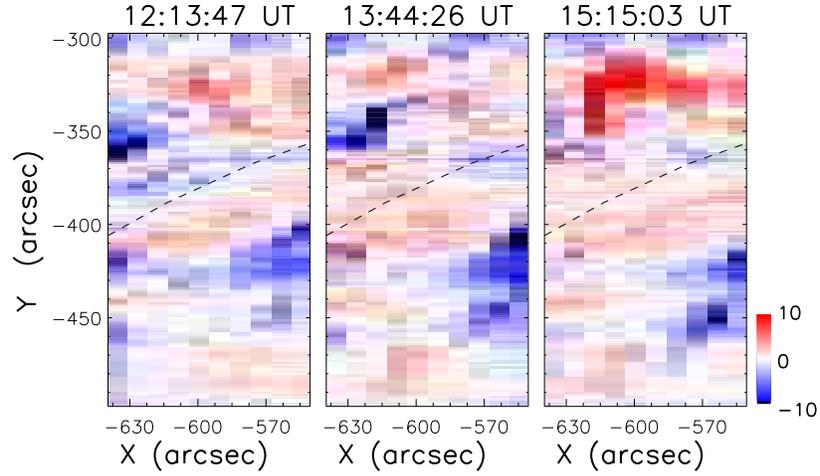}
\caption{Time evolution of the \ion{Fe}{12} 195 \AA\ Dopplergrams, where blue color represents blue shift and red color represents red shift. The dashed lines are the magnetic polarity inversion line. The colour bar shows the velocity range in units of km s$^{-1}$. }
\label{fig6}
\end{figure}

The {\it Hinode}/EIS observations started at 12:02:37 UT, which was after the rise of the filament but before the flaring ribbons appeared. Figure \ref{fig6} displays the \ion{Fe}{12} 195 \AA\ Dopplergrams at four times, where the dashed line marks the location of the magnetic PIL. It is seen that the area to the south of the PIL, i.e., in the leading positive polarity, consists of patches of both blue and red shifts, with the blue shift more prominent. The corresponding upflows had a typical velocity of 1--10 km s$^{-1}$. Despite the dominance of the blue shift, weak red shift with a velocity of 1--4 km s$^{-1}$, which is around the error of the line-profile fitting, was also present.

To the north of the magnetic PIL, the field of view (FOV) of EIS is also mixed with red and blue shifts, with the red shift dominant, however. The downflow velocity was up to 10 km s$^{-1}$ near the top of the FOV of EIS. The strong red-shifted channel was bracketed by two narrow blue-shifted strips, which were persistent. However, the strong blue-shifted patch around (-630\arcsec, -350\arcsec) was present before 15:00 UT, and disappeared afterwards. It was probably associated with the erupting filament.

\section{Discussions}\label{sec4}

\subsection{Chirality and Magnetic Configuration of the Filament}

Barbs are fine structures of solar filaments. Seen from above, each barb makes an acute angle with respect to one direction of the filament spine. When viewed along this direction, the barb is located either on the left
side or the right side of the spine. Based on this difference, filaments can be classified into either the left-bearing or the right-bearing types \citep{mart92}. On the other hand, judged from the direction of the internal magnetic field, filaments can be divided into sinistral and dextral types, where the internal magnetic field, when viewed from the positive polarity of the filament channel, is pointed to the left/right for the sinistral/dextral filaments. Such a chirality is directly related to the magnetic helicity, i.e., a sinistral filament has positive helicity, whereas a dextral filament has negative helicity. It was found that there is a one-to-one correspondence between left-bearing/right-bearing filaments and the sinistral/dextral chirality \citep{mart92}. However, recently \citet{guo10} found that this rule is not universal. With the nonlinear force-free magnetic field extrapolation, they found that there are both left- and right-bearing barbs in different portions of a filament. The chirality rule is applicable to the portion of the filament embedded in a flux rope, and is opposite in another portion of the filament whose magnetic field is sheared arcades.

\begin{figure}
\epsscale{0.8}
\plotone{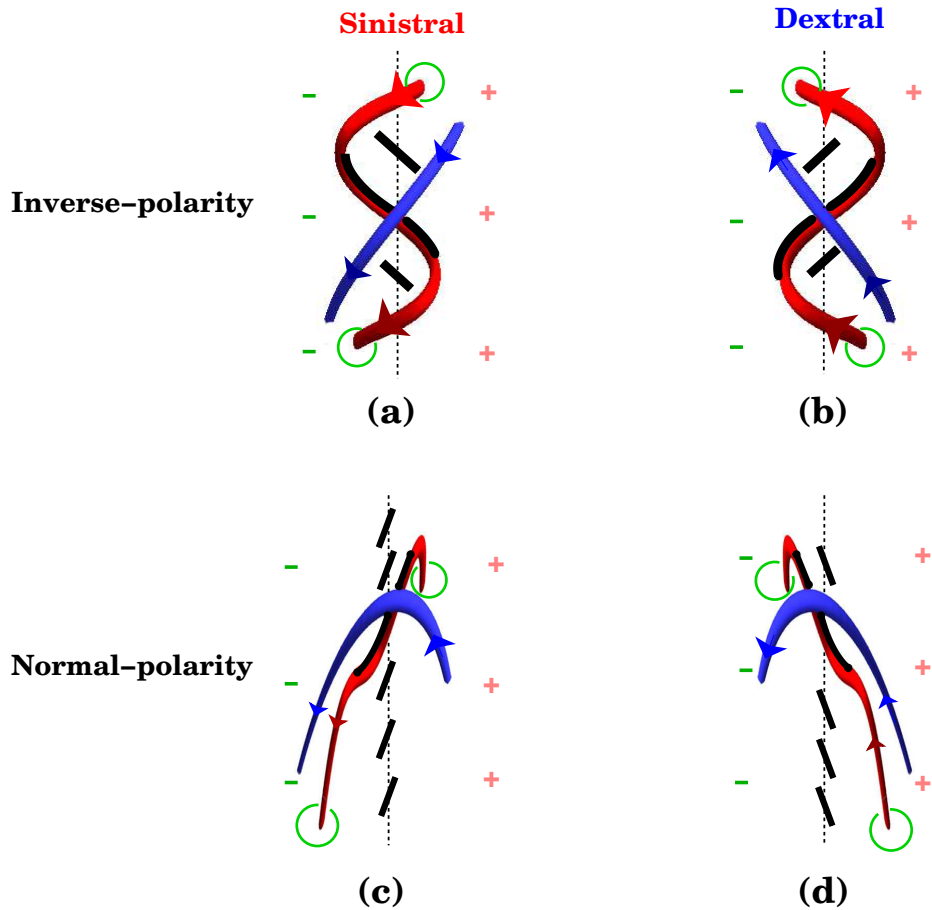}
\caption{Schematic sketch illustrating the relationship between the chilarity of a filament and the orientation of the threads and barbs in the inverse-polarity magnetic configuration ({\it top}) and the normal-polarity configuration ({\it bottom}). The dashed lines are the magnetic PIL, the red lines represent the core magnetic field, the thick black lines represent the filament threads hosted by the dips of the core field, and the blue lines represent the envelop magnetic field. The green circles mark the locations of the filament drainage when eruption happens. Panel (a):  left-bearing barbs and right-skewed coronal arcades; Panel (b): right-bearing barbs and left-skewed arcades; Panel (c): right-bearing barbs and right-skewed arcades; Panel (d): left-bearing barbs and left-skewed arcades.}
\label{fig7}
\end{figure}

Observations indicate that filaments can be divided into inverse- and normal-polarity types, depending on whether the magnetic field component perpendicular to the filament spine is the same as or opposite to what expected from the photospheric polarities. \citet{kr74} proposed a flux rope for the inverse-polarity filaments (henceforth abbreviated to ``K-R model"), where the flux rope is detached from the solar surface; on the other hand, \citet{ks57} proposed a dipped arcade model for the normal-polarity filaments (henceforth abbreviated to ``K-S model"). These models are based on two-dimensional considerations. In three dimensions, the K-R model has been extended into a weakly twisted ($\sim$1.5 turns in most cases) flux rope, whose footpoints are anchored to the solar surface \citep[e.g.,][]{aula98, vanb04}, and the K-S model has been extended into the sheared arcade model \citep[e.g.,][]{anti94}. From a theoretical point of view, the three-dimensional extension of the K-R and K-S models can be divided into four categories in terms of sinistral and dextral chiralities, which are illustrated by the four panels in Figure \ref{fig7}. In this figure, the dashed lines are the magnetic PIL; the red lines represent the twisted or strongly sheared core field whose dips support the filament threads, which are shown as black lines; the blues lines represent the less sheared envelop field. For each case, only one line in the core magnetic field and another line in the envelop magnetic field are plotted for conciseness. The drawn core field line has a longer magnetic dip than other core field lines, hence it has a longer thread. The longer thread would be observed as a barb when viewed above (Note that the model that a barb can be represented by a longer thread is just one scenario, which is backed by the fact that many barbs are not located above the local PIL associated with a parasitic polarity as found by \citealt{liu10}. When associated with a parasitic polarity, a barb can also be a collection of threads as mentioned by \citealt{chae05}). As a result, a sinistral inverse-polarity filament has left-bearing barbs and right-skewed coronal arcades (panel a); a dextral inverse-polarity filament has right-bearing barbs and left-skewed coronal arcades (panel b); a sinistral normal-polarity filament has right-bearing barbs and right-skewed coronal arcades (panel c); a dextral normal-polarity filament has left-bearing barbs and left-skewed coronal arcades (panel d). From this diagram we can see that only in the flux rope model which describes the inverse-polarity filaments, there is a one-to-one correspondence between the left-/right-bearing filaments and the sinistral/dextral chirality as found by \citet{mart92}. For the sheared arcade model which describes the normal-polarity filaments, the correspondence is opposite, i.e., a left-bearing filament has a dextral chirality and a right-bearing filament has a sinistral chirality, as found by \citet{guo10}. The original chirality rule found by \citet{mart92} might be due to the prevalence of inverse-polarity filaments in the solar atmosphere \citep{lero84}.

When a filament erupts, the hosting magnetic field would rise and expand. As a result, part of the filament material would fall down to the solar surface along the two legs of the magnetic flux tube under the frozen-in condition. If the theoretical models presented in Figure \ref{fig7} are correct, we see that the two conjugate footpoints of the filament drainage as indicated by the green circles in Figure \ref{fig7}, when connected, should have the same skewness as the flaring arcade \citep{mcal97}, i.e., for sinistral filaments, the two conjugate draining sites are right-skewed, whereas for dextral filaments, the two conjugate draining sites are left-skewed. For the filament eruption event studied in this paper, the southern draining site of the filament was located on the east side of the magnetic PIL, as indicated by Figure \ref{fig2}. Therefore, it can be claimed that the two conjugate draining sites are right-skewed, which is exactly the same as the flaring arcade as indicated by Figure \ref{fig5}(b), implying that the filament has a sinistral chirality.

\begin{figure}
\epsscale{0.7}
\plotone{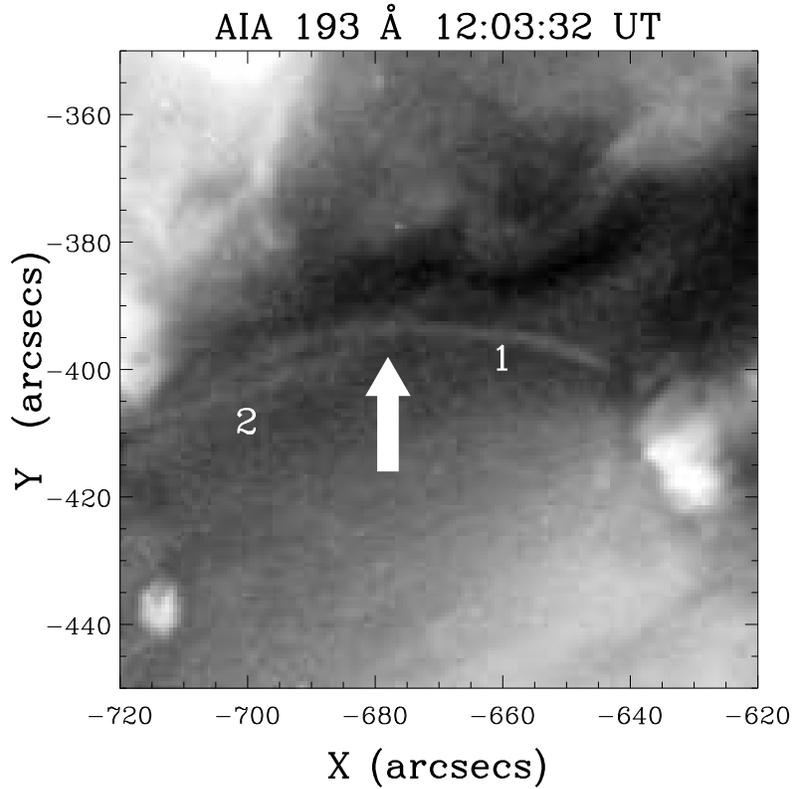}
\caption{A snapshot of the AIA 193 \AA\ map at 12:03:32 UT showing the brightening of two interweaved strands which are labeled as ``1" and ``2". One of their crossings is marked by the white arrow. It is seen that strand 1 is above strand 2 at this crossing, implying a positive mutual helicity of the magnetic system.}
\label{fig8}
\end{figure}

The sinistral chirality of this filament can be examined by an independent method proposed by \citet{chae00}, who used the crossing of two threads in a filament to determine the magnetic helicity of the filament, i.e., for two threads at different altitudes crossing each other apparently in the image, if the upper thread is right-skewed relative to the lower thread, the filament has a positive helicity; if the upper thread is left-skewed relative to the lower thread, the filament has a negative helicity. The AIA 193 \AA\ image of the erupting filament at 12:03:32 UT is shown in Figure \ref{fig8}, where two bright strands beside the filament threads, labeled ``1" and ``2", crossed each other twice. At the crossing indicated by the white arrow, strand ``1" was in the foreground and therefore was above strand ``2". Since strand ``1" is right-skewed relative to strand ``2", the mutual helicity should be positive, which is consistent with the sinistral chirality determined by the skewness of the filament drainage.

Once the chirality of an erupting filament is determined with the methods above, we can deduce the magnetic configuration of the filament by combining the bearing sense of the barbs (or longer threads) based on the diagram in Figure \ref{fig7}.  For a sinistral filament, if the barbs have left-bearing, the filament should be of the inverse-polarity type; if the barbs have right-bearing, the filament should be of the normal-polarity type. For a dextral filament, the correspondence is opposite. Taking the erupting filament in this paper as an example, the inset in Figure \ref{fig3}(a) indicates that the filament has left-bearing barbs. Therefore, we can conclude that this portion of the filament is of the inverse-polarity type. Note that magnetic dips with normal-polarity and inverse-polarity might co-exist in one filament sometimes,
as demonstrated by \citet{aula02}.

\subsection{The Nature of H$\alpha$ Counter-streamings}

Counter-streamings discovered by \citet{zirk98} are believed to exist in every filament \citep{deng02, lin03, schm08}. Therefore, although there is no H$\alpha$ Doppler observation nor high-resolution H$\alpha$ filtergrams for the filament under study in this paper, we presume that counter-streamings exist in this filament as well. The natural explanation for the counter-streamings is that they result from the longitudinal oscillations of threads confined near the magnetic dip, where the oscillations are not in phase with each other. The oscillatory nature of the H$\alpha$ counter-streamings was verified by \citet{ahn10}, who traced the motion of the streaming threads and found that some of them moved to an apex and then turned back. Such oscillations have been simulated by \citet{anti00}, \citet{xia11}, \citet{luna12}, and \citet{zhan12,zhan13}. It is noted that actually there exists another possibility, i.e., each thread presents a dynamic equilibrium along each flux tube. The hot coronal plasma siphons into the H$\alpha$ filament thread from one end, where it cools down and keeps moving with a lower velocity in order to maintain the mass flux. At the other end of the H$\alpha$ thread, the plasma siphons out into the coronal portion of the flux tube, and finally drains down to the solar chromosphere. Such a dynamic equilibrium of an apparently unchanging filament thread has been well studied even before counter-streaming was discovered \citep[e.g.,][]{tsin92, delz96}. Since each flux tube and the embedded H$\alpha$ thread are independent from others, the moving directions of neighboring flux tubes might be random, especially if the siphon flow is driven by the random convective motions at the footpoints of the flux tube. As a result, we may see blue- and red-shifted counter-streamings inside a filament. If this mechanism works, we expect to see that the assembly of the legs of the flux tubes should also be prevailed by counter-streamings. However, the EUV Dopplergram in Figure \ref{fig6} indicates that there is no signature of intermittent upward and downward velocity field in the small spatial scale as in H$\alpha$ Dopplergrams shown by previous researchers, e.g., \citet{zirk98} and \citet{lin03}. Instead, the EUV Dopplergrams near the magnetic PIL reveal large patches of either blue shift or red shift with a size of tens of arcseconds. It might be argued that the horizontal step of the {\it Hinode}/EIS raster observation, 10\arcsec, is too large to resolve the small-scale counter-streamings. However, even along the slit direction, which has a pixel size of 1\arcsec, we still cannot see this type of counter-streamings.

These observations are similar to those in \citet{engv85}, who found that the filaments were located very close to the dividing lines between areas of upflows and downflows. Here, with higher spatial resolution, we find that even on one side of the filament, there are patches of upflows and downflows. Within the FOV of {\it Hinode}/EIS, both the northern and the southern sides of the magnetic PIL were mixed with upflow and downflow patches. Therefore, while there is no EUV counterpart of the fine-structured H$\alpha$ counter-streaming along the filament channel implied by previous researchers, e.g., \citet{zirk98} and \citet{lin03}, there exists a different type of large-scale counter-streaming in the EUV Dopplergram. At the legs of the magnetic flux tubes along a filament channel, upflows or downflows with a velocity of up to $\sim$10 km s$^{-1}$ cluster in patches with a size of tens of arcseconds or larger. The time evolution of the EUV Dopplergram indicates that such patchy upflow/downflow motions are persistent, with slight variations. It is noted that the Doppler velocity of the filament channel in \ion{C}{4} 1548 \AA\ and \ion{Si}{4} 1393 \AA\ ($\log T \sim 10^5$ K) presented by \citet{engv85} was in the range of $\pm15$ km s$^{-1}$, which is larger than our value for the \ion{Fe}{12} 195 \AA\ hotter line. Since the spectral profiles appeared like single Gaussian we fitted the profiles in such a way which gives the bulk speed of plasma. There has been other work looking at multiple profile fitting which would lead to some larger velocities \citet[e.g.,][]{tian12} which would exist for weaker plasma. 

The physical meaning of this result is two-fold. First, the absence of counter-streaming with a spatial scale as small as in Figure 2 of \citet{lin03} means that the H$\alpha$ counter-streamings, which were found to be ubiquitous in filaments \citep{zirk98, lin03}, are mainly due to the longitudinal oscillations of filament threads. Threads in neighboring flux tubes do not oscillate in phase, resulting in the effect of counter-streamings. On the other hand, the persistent upflow or downflow motions at the legs of the flux tubes imply that there is also another component of a uni-directional motion for a cluster of flux tubes as modeled by \citet{low05} and \citet{petr05}. Such a large-scale counter-streaming motion is similar to the antiparallel streaming inside filaments as revealed by {\it Hi-C} mission \citep{alex13}. The real motion of the H$\alpha$ counter-streamings found in previous filament observations is a superposition of an oscillation and a uni-directional flow. This explains why the measured velocity of individual H$\alpha$ threads in \citet{lin03} was oscillating around a limited  value (not zero). As simulated by \citet{xia11}, if the magnetic dip is too shallow, the filament thread will drain down to one footpoint of the hosting flux tube; if the magnetic dip is deep, the filament thread keeps oscillating, with the central position deviating from the trough of the magnetic dip. 

\begin{figure}
\epsscale{0.7}
\plotone{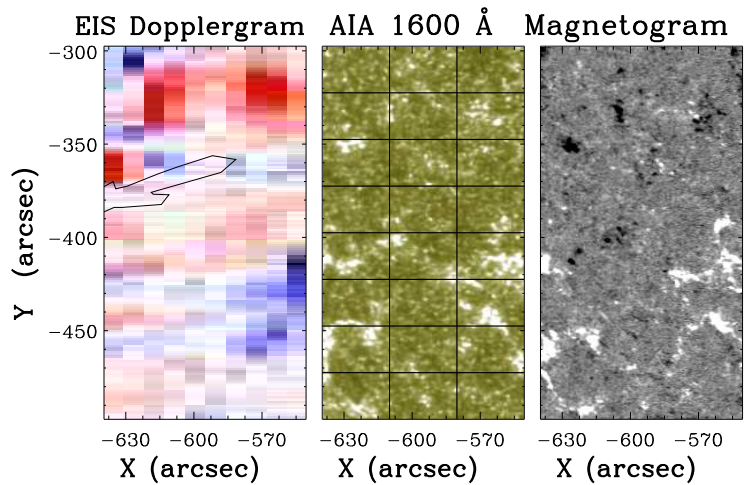}
\caption{Comparison of the \ion{Fe}{12} Dopplergram at 16:57:01 UT ({\it left}), the AIA 1600 \AA\ intensity at 12:02:17 UT ({\it middle}), and the HMI magnetogram at 16:57:25 UT ({\it right}) within the field of view of {\it Hinode}/EIS. The black line in the left panel outlines the location of the H$\alpha$ filament well before the eruption, the grid in the middle panel divides the field of view into $3\times 8$ sub-areas.}
\label{fig9}
\end{figure}

\begin{figure}
\epsscale{0.7}
\plotone{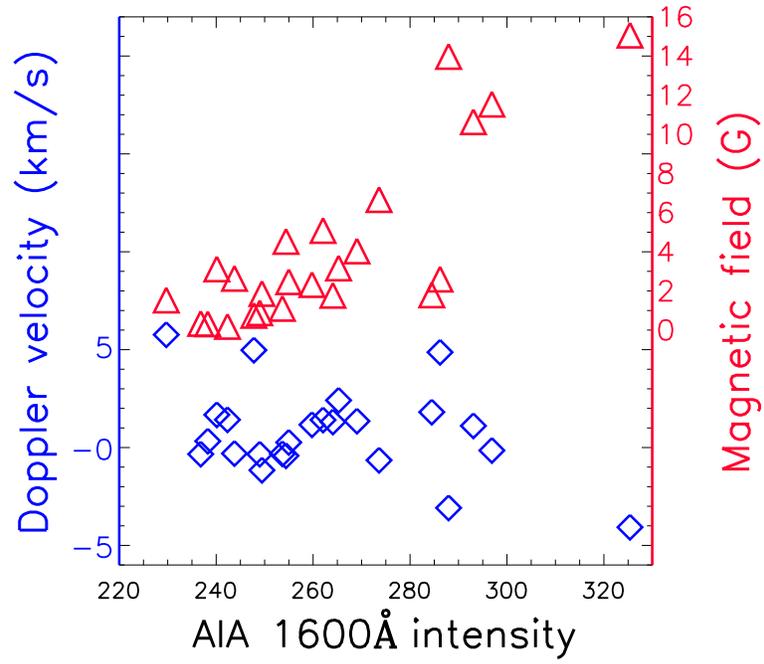}
\caption{Scatter plots of the correlation between the \ion{Fe}{12} 195 \AA\ Doppler velocity and the AIA 1600 \AA\ intensity ({\it blue diamonds}) and the correlation between the magnetic field and the AIA 1600 \AA\ intensity ({\it red triangles}) in the $3\times 8$ sub-areas shown in Fig. \ref{fig9}.  Negative/positive values of the Doppler velocity mean upflow/downflow.}
\label{fig10}
\end{figure}

A further question is what drives the upflows/downflows along the filament channel and the uni-directional flow along each flux tube. In the one-dimensional simulations \citep{anti00, karp06, xia11}, it is generally assumed that there is additional heating localized in the chromosphere. In order to investigate the cause of the upflow and the downflow along the filament channel, we compare one snapshot of the \ion{Fe}{12} Dopplergram, the AIA 1600 \AA\ intensity, and the HMI magnetogram in Figure \ref{fig9}, where the location of the H$\alpha$ filament at 03:00:14 UT is marked by the black line in the left panel, and the Dopplergram is chosen at 16:57:01 UT when the erupting filament was probably not in the FOV of the {\it Hinode}/EIS. The AIA 1600 \AA\ intensity map, which is corrected with the limb-darkening effect, reveals that the filament was surrounded by bright patches on both sides with the positive and negative polarities, which are active networks (Note that active networks are decayed plages. They are visible in AIA 1700 \AA, but invisible in AIA 4500 \AA\ white-light band). Based on the sinistral chirality of the filament, the southern side of the PIL, i.e., the positive polarity, should be magnetically linked to the east, which is outside the FOV of {\it Hinode}/EIS, and the northern side of the PIL, i.e., the negative polarity, should be linked to the west of the FOV of {\it Hinode}/EIS. Therefore, limited to the small FOV, we cannot compare the velocity pattern of two conjugate footpoints of flux tubes. However, we can still get the impression from Figure \ref{fig9} that the upflow, i.e., blue shift, is related to the  brighter portion of the chromospheric active networks and the stronger photospheric magnetic field. Since the \ion{Fe}{12} line is formed in the corona, the AIA 1600 \AA\ emission comes from the temperature minimum region and the transition region, whereas the magnetogram is measured in the photosphere, therefore we cannot compare their values pixel by pixel. Instead, we divide the FOV of the {\it Hinode}/EIS into $3\times 8$ sub-areas as illustrated in the middle panel of Figure \ref{fig9}, and then take the average values of the AIA 1600 \AA\ intensity, Doppler velocity, and the magnetic field strength in each sub-area. The correlations between the former and the latter two parameters are displayed in Figure \ref{fig10} as scatter points. Without doubt there is a strong correlation between the 1600 \AA\ intensity and the local magnetic field, as revealed by the red triangles \citep[see also][]{louk09}, with the correlation coefficient being 0.82. Albeit weak, a tendency can also be discerned that upflows (with a negative velocity) are related to the brighter 1600 \AA\ intensity, as revealed by the blue diamonds. Their correlation coefficient is -0.37, suffering from the  displacement between the two wavebands. After making the moving correlation analysis, we find that the correlation coefficient between the Doppler velocity and the AIA 1600 \AA\ intensity can be as high as -0.69 when the FOV of the sub-region in the AIA image is shifted by $\Delta x$=4\arcsec\ and $\Delta y$=17\arcsec. Such shifts with positive values are consistent with the projection effects since the EUV emissions are from a higher altitude and the region of interest was located in the southeast quadrant of the solar disk. The larger $\Delta y$ in contrast to smaller $\Delta x$ implies that the legs of the magnetic loops deviate from the local radial direction. Therefore, we can conclude that the persistent flow in the filament channel is driven by the extra heating in the lower atmosphere of the decayed plage, which is magnetic in nature \citep{dela13}.

The dynamics of filament threads can be conjectured as follows: For flux tubes rooted in the plages or active networks, which are presumably heated by Alfv\'en waves, the unequal heating at the two footpoints drives a siphon flow along the flux tube with enhanced density in the corona. If the heat imbalance between the two footpoint is suitable (not too large), the plasma density inside the magnetic loop satisfies the criterion of thermal instability so that a filament thread is formed, as simulated by \citet{anti00} and \citet{xia11}. Once formed, the filament thread begins to oscillate around the magnetic dip under the perturbation of the persistent flow. At the same time, since the persistent existence of the uni-directional flow, hot plasmas keep penetrating into the thread from one end and moving out from the other end, i.e., the measured velocity of a filament thread consists of two parts, the oscillation and the uni-directional flow. When the uni-directional flow is strong enough, the thread may drain down toward the footpoint of the flux tube with weaker heating.

Since H$\alpha$ counter-streamings seem to be the thread oscillations superposed on a uni-directional flow according to the paradigm presented above, the filament counter-streamings are essentially the same as the longitudinal oscillations of filaments as discovered by \citet{jing03} with the only difference being whether the neighboring threads are oscillating in phase or not. Since different flux tubes may have different curvature, hence the thread oscillations will have different periods \citep{luna12, zhan13}. Therefore, even a collection of threads may oscillate in phase initially when they are perturbed, they will finally be out of phase. With that, filament longitudinal oscillations turn to be counter-streamings. In this sense, the enhanced counter-streamings as a possible precursor of CMEs found by \citet{schm08} can be unified with the filament oscillations as a precursor of CMEs found by \citet{chen08}.

\subsection{Reconnection Rate of the Associated Flare}

The first solar flare discovered more than 150 years ago was a white-light flare, where the white-light emission is from the lower chromosphere and photosphere. It is now generally believed that solar flares are due to magnetic reconnection in the corona, where the heat conduction and/or nonthermal particles transfer energy down to heat the chromosphere. When the chromospheric emission is strong, it may further heat the photosphere through back-warming \citep[see][and references therein]{cheng10}. For most flares, no discernable white-light brightening can be detected, and the flaring emissions are mainly from the corona in the X-ray and EUV wavelengths and from the chromosphere in the optical lines and UV wavelengths.

For the filament eruption in this paper, we can see that the flaring loops and the two ribbons in EUV for more than 10 hrs, from the hot line (211 \AA\ at 2 MK) to the warm line (304 \AA\ at 0.05 MK). However, there was no any signature of ribbons or brightenings in chromospheric H$\alpha$. By measuring the velocity ($v$) of the separation of the flare ribbons in EUV, which is then multiplied by the photospheric magnetic field ($B$), we can estimate the reconnection rate in terms of the electric field strength in the reconnection site as proposed by \citet{forb84}. The velocity of the northeastern ribbon is $v\sim 0.7$ km s$^{-1}$, and the photospheric magnetic field is of the order of 1 G, the resulting reconnection rate is therefore 0.07 V m$^{-1}$. This value is 3--4 orders of magnitude smaller than those in M- and X-class flares \citep{qiu04}. Presumably because the reconnection rate of this long-duration flaring event is so low that the weak heating from the coronal magnetic reconnection is transferred down to the transition region only.

\section{Summary} \label{sec5}

In this paper, we investigate a filament eruption event occuring on 2011 January 29, with the help of EUV spectroscopic observation from {\it Hinode}/EIS and imaging observations from {\it GONG}/H$\alpha$, {\it STEREO}/EUVI, {\it SDO}/AIA and {\it SDO}/HMI. The main results are summarized as follows.

(1) Erupting filaments are frequently observed to drain part of the plasma along two legs down to the solar surface. We propose that the connected two conjugate sites of drainage on the solar surface have the same skewness as the magnetic field or flaring arcade. Therefore, right-skewed drainage sites correspond to a sinistral filament, whereas left-skewed drainage sites correspond to a dextral filament. It is also suggested that, combined with the bearing of the filament barbs (left or right), this rule can be used to judge whether a filament has inverse polarity or normal polarity.

(2) We confirm that the observed H$\alpha$ counter-streamings in solar filaments are the longitudinal oscillations of filament thread in nature. This type of small-scale counter-streaming is absent in the coronal portion of the filament channel. However, it is revealed that there are large-scale patchy ``counter-streamings" in the coronal portion of the filament channel. Within each patch, there is uni-directional flow from one magnetic polarity to the other through the filament thread. We showed evidence that the direction of the flow is determined by the inequality of the magnetic field of the two footpoints. The upflow comes from the footpoint with stronger magnetic field and hence brighter active network or plage. As a result, the dynamics of filament threads is a superposition of longitudinal oscillations and a uni-directional flow.

(3) The associated long-duration (more than 10 hrs) two-ribbon flare is found to be discernable in EUV only, leaving no any signature in the chromosphere. This is probably due to its too low reconnection rate, 0.07 V m$^{-1}$, which is about 4 orders of magnitude smaller than those of M- and X-class flares.

\acknowledgments
The authors thank the referee for constructive comments and the {\it SDO}/AIA, {\it STEREO}/EUVI, and {\it GONG} teams for providing the data. Hinode is a Japanese mission developed and launched by ISAS/JAXA, collaborating with NAOJ as a domestic partner and NASA and STFC (UK) as international partners.{\it SOHO} is a project of international cooperation between ESA and NASA. This research is supported by the Chinese foundations 2011CB811402 and NSFC (11025314,
10933003, and 10673004).

\clearpage

\end{document}